# Optical manipulation with metamaterial structures

Yuzhi Shi[1,2,3,4], Qinghua Song[5]*, Ivan Toftul[6,7], Tongtong Zhu[8], Yefeng Yu[9]*, Weiming Zhu[10]*, Din Ping Tsai[11], Yuri Kivshar[6,7]*, and Ai Qun Liu[12]*

[1]MOE Key Laboratory of Advanced Micro-Structured Materials, Shanghai 200092, China.

[2]Institute of Precision Optical Engineering, School of Physics Science and Engineering, Tongji University, Shanghai 200092, China

[3]Shanghai Institute of Intelligent Science and Technology, Tongji University, Shanghai 200092, China

[4]Shanghai Frontiers Science Center of Digital Optics, Shanghai 200092, China.

[5]Tsinghua Shenzhen International Graduate School, Tsinghua University, Shenzhen, 518055, China

[6]Nonlinear Physics Centre, Research School of Physics, Australian National University, Canberra ACT 2601, Australia

[7]School of Physics and Engineering, ITMO University, St. Petersburg 197101, Russia

[8]School of Optoelectronic Engineering and Instrumentation Science, Dalian University of Technology, Dalian, 116024, China

[9]School of Electronic and Optical Engineering, Nanjing University of Science and Technology, Nanjing 210094, China

[10]School of Optoelectronic Science and Engineering, University of Electronic Science and Technology of China, Chengdu 610054, China

[11]Department of Electrical Engineering, City University of Hong Kong, Tat Chee Avenue, Kowloon, Hong Kong, China

[12]School of Electrical and Electronic Engineering, Nanyang Technological University, Singapore 639798, Singapore

*e-mail: song.qinghua@sz.tsinghua.edu.cn (Q.S.); zhuweiming@uestc.edu.cn (W.Z.); yf_yu@njust.edu.cn (Y.Y.); yuri.kivshar@anu.edu.au (Y.K.); eaqliu@ntu.edu.sg (A.Q.L.)

# ABSTRACT

Optical tweezers employing forces produced by light underpin important manipulation tools in many areas of applied and biological physics. Conventional optical tweezers are based on refractive optics, and they require excessive auxiliary optical elements to reshape both amplitude and phase, as well as wavevector and angular momentum of light, and thus impose limitations to the overall cost and integration of optical systems. *Metamaterials* provide both electric and optically induced magnetic response in subwavelength optical structures, and they are highly beneficial to achieve unprecedented control of light required for many applications, also opening new opportunities for *optical manipulation*. Here, we review the recent advances in the field of optical tweezers employing the physics and concepts of metamaterials (the so-called "meta-tweezers") and demonstrate that metamaterial structures could not only advance classical operations with particles, such as trapping, transporting, and sorting, but they uncover *exotic optical forces* such as pulling and lateral forces. Remarkably, apart from manipulation of particles, metastructures can be powered dynamically by light to realize ingenious "meta-robots". We provide an outlook for future opportunities in this area ranging from enhanced particle manipulation to meta-robot actuation.

**TABLE OF CONTENTS**

# I. INTRODUCTION

Ever since the pioneer experiments performed by Arthur Ashkin and his collaborators, optical tweezers have revolutionized the way of manipulating tiny objects (such as viruses, DNA, and proteins),[1-5] and have benefited a diversity of applications in many disciplines, including nanotechnology, biology, and quantum sciences.[6-10] Optical tweezers based on refractive optics range from single-beam to multi-beam geometries, and they have been widely utilized to trap single viruses and proteins,[11] stretch DNA,[12] and manipulate cold atoms.[13] Spatially structured beams produced by stacked optical elements or spatial light modulator (SLM) can reshape the amplitude and phase of light, thus offering many possibilities of particle manipulations, for instance, rotation by Laguerre-Gaussian (LG) or elliptically polarized beams,[14-16] sorting in interference patterns[17,18] using microlens[19,20] or holography,[21,22] assembly and cooling of atoms in optical lattices,[13,23-25] three-dimensional clearing with Airy beams.[3,26-28] Unconventional steering of beam vector provides extra degrees of freedom in dynamic manipulation, far-reaching examples include the pulling of particles in the Bessel-like beams,[29-33] lateral forces acting on chiral particles from helix beams,[34-38] as well as the torque control in topological light fields.[39,40] Elliptically polarized light carries spin angular momentum (SAM), and it can spin polarization-anisotropic particles, whereas vortex beams with orbital angular momentum (OAM) exert orbital forces to rotate particles.[41] Intriguingly, single circularly polarized focused beams were demonstrated to rotate particles in orbit by employing spin-orbit interactions (SOI).[42-45] Recently, the extraordinary transverse spin, which is orthogonal to the wavevector, has attracted a great deal of attention after it was reported theoretically for focused light beams,[46] evanescent[45] and interfering waves,[47] and it was demonstrated experimentally.[48-52] Despite enormous efforts and manifold applications, the free-space optical tweezers are limited because they use bulky optical systems, have diffraction-limited resolutions, and have low efficiency for the manipulation of large quantities of nanoobjects (e.g., viruses and proteins).

Near-field optical tweezers based on nanowaveguides, photonic crystal cavities, and nanoprobes push the limit of optical manipulation towards smaller sizes, higher efficiency, and larger quantity of trapped objects by virtue of their nontrivial capabilities in localizing light fields. For example, a slot waveguide is capable to trap and transport single RNA molecules in microfluidic environment.[53,54] Photonic-crystal structures can trap nanoparticles and biomolecules by employing cavities or waveguides.[55-59] Plasmonic nanoantennas use the highly localized fields enabled by surface plasmons to trap nanoparticles and biomolecules with sizes < 20 nm.[2,60-62] Meanwhile, the near-field nanoprobe is able to extract single DNA directly from a cell,[63] or trap a cell at the tip for subdiffractional imaging.[64] Though the above techniques in near-field photonics have already advanced optical tweezers in many ways, e.g. for massive sorting of nanoparticles with different sizes[65] or shapes,[66] dynamic patterning of nanoparticles,[2,57,62,67] for the study of counterintuitive light-matter interactions,[34,35,38,45] the optical manipulation focuses mainly on enhancing the light intensity rather than controlling its phase, as well as the exploration of interaction between a light wave and special particles (e.g., chiral) rather than reconstructing the momentum topology. It is worth noting that, our review paper has a special focus on metamaterial structures, whereas comprehensive reviews of optical tweezers and their applications using classical optics can be found elsewhere.[2,3,15,29,54,62,68]

Metamaterials,[69-74] or their two-dimensional relatives – metasurfaces,[75-80] are artificially designed architectures composed of subwavelength elements. They exhibit many unique properties that are rarely or never found in nature, and they have unprecedented advantages to reshape the light fields, thus creating a distinctive paradigm for optical manipulation. The interactions between light and nanostructures can easily modulate the phase and amplitude of the wavefront,[81-88] freely control the directions of wave vectors,[89-91] arbitrarily change the beam spatiotemporal dynamics,[92-96] and harnesses efficiently the angular momentum of light.[97,98] Here, we review the recent advances of optical manipulation utilizing metamaterials for optical tweezers. We show that, by shaping the amplitude and phase of light, and employing magnetic response of metastructures, we can create "meta-tweezers" for miniaturizing optical systems and enhancing optical forces via distinct new emerging physics (see Fig. 1); by engineering the momentum topology of light, "meta-tweezers" can enable counterintuitive optical pulling forces, binding, etc; by dynamically controlling the spatial and temporal distributions of forces and potential wells, as well as the SAM or OAM of light, particles can be dynamically positioned, and the novel "meta-robot" with mobilized meta-nanostructures can be realized, paving a new avenue for the designing of light powered nano-micro robots, which could have enormous biophysical applications.

In this paper, first we outline the fundamental physics behind optical forces and metamaterials directly related to the concepts of optical manipulation using "meta-tweezers" and "meta-robots". More specifically, in Sec. II we discuss the methods for calculating optical forces including dipole forces, angular momentum-dependent forces, phase-dependent forces, and wavevector-dependent forces. We also provide the explicit expression for an optical force acting on a dipole particle placed near a surface, that underpins the most common scenario of optical manipulations employing metasurfaces. The fundamental physics of metamaterials for manipulating the amplitude, phase, wavevector, angular momentum of light, and other degrees of freedom are summarized in Sec. III. Section IV is the central section of our paper, and it presents several specific examples of optical manipulation with metamaterial structures including metasurfaces. Finally, Section V concludes the paper with perspectives and outlines.

## II. A SUMMARY OF OPTICAL FORCES
### A. General approach

Light carries linear momenta which can be transferred to matter. Considering an arbitrary particle in the field of incident electromagnetic field, the momentum of the light tends to change during the scattering process. The value of the momentum change is exactly the momenta that particle gained with different sign due to the momentum conservation law, which will be quantitatively analyzed in the following. Since most experiments on optical tweezers are utilizing lasers, we limit ourselves in this review to only monochromatic electromagnetic fields with frequency $\omega$. Most formulas can be written in a compact form using complex amplitudes $\mathbf{E}$, $\mathbf{H}$ of the real observable fields $\mathcal{E}$, $\mathcal{H}$, which are connected as $\mathcal{E}(\mathbf{r},t) = \text{Re}\left[\mathbf{E}(\mathbf{r})e^{-i\omega t}\right]$ and $\mathcal{H}(\mathbf{r},t) = \text{Re}\left[\mathbf{H}(\mathbf{r})e^{-i\omega t}\right]$, where $\mathbf{r}$ and $t$ represent the location and time, respectively.

Rigorous calculation of optical forces on arbitrary sized and shaped particles can be performed by calculating the average change in momentum flux via scattering. Technically, it is done by integrating the flow of Maxwell stress tensor through a closed surface $A$ which contains the scatterer[54]

$$\langle \mathbf{F} \rangle = \oint_A \langle \overleftrightarrow{\mathbf{T}} \rangle \cdot \hat{\mathbf{n}} \, \mathrm{d}A . \tag{1}$$

Here, $\hat{n}$ is the unit vector outward normal to the integral surface $A$ which to be noted can be arbitrary, the symbol $\langle \cdot \rangle$ means the time average over an oscillation period. The Minkowski form of the stress tensor can be given as

$$\langle T_{ij} \rangle = \frac{1}{8\pi} \text{Re}\left[ D_i E_j^* + B_i H_j^* - \frac{1}{2}\left( \mathbf{D} \cdot \mathbf{E}^* + \mathbf{B} \cdot \mathbf{H}^* \right)\delta_{ij} \right]. \tag{2}$$

where $\delta_{ij}$ is Kronecker delta, $\mathbf{D} = \varepsilon\mathbf{E}$ and $\mathbf{B} = \mu\mathbf{H}$, where $\varepsilon$ and $\mu$ are the electric permittivity and magnetic permeability of the host medium, respectively. Other forms of stress tensor include those by Abraham, Lorentz, as well as by Einstein and Laub can be found elsewhere,[99] while the Minkowski form is generally accepted in the community of optical forces. It is sometimes crucial when one calculates force on a floating object.[100] The relavent well-known Abraham–Minkowski controversy in electrodynamics can be found in detail in Refs.[100-103]. We will not cover in complete details the controversy but give the general quick overview of it. This long-lasting debate is mainly on how to calculate momentum of light inside medium. Initially, it was started by Abraham and Minkowski in their classical works.[104,105] Fundamentally, optical force and torque are caused by the change of linear or angular momenta of scattered photons. A major part of main pioneer experiments on optical manipulation were performed with suspended particles in fluids. The well established theory for the free space[106] met obstacles when question "what is the momentum of a photon in a bulk medium?" had been raised.[107-110] The situation gained much in complexity when the effects of dispersive medium came into the play[109,111-113] and even dispersive bianisotropic medium.[114]

For the numerical calculations, Equation (1) is generally used, while the integration boundary $A$ can be arbitrary, the better numerical accuracy gives auxiliary boundary with a slightly bigger radius than the particle as the mesh is much denser in the near field region.[115]

### B. Optical forces acting on a dipole

A particle can be regarded as a *dipole* when the typical radius satisfies $a \ll \lambda$, where $\lambda$ is the wavelength in the host medium with a refractive index of $n_m$. The common dimensionless parameter here is the so-called size parameter $x = ka = \dfrac{n_m \omega}{c} a$, where $k$ is the wave number in the medium and $c$ is the light velocity in free space.

Sufficiently, small particles can support only first multipoles: electric dipole **p** and magnetic dipole **m**. Those dipoles are induced by the incident field and linearly depends on electric and magnetic fields with the proportionality factor called polarizability, defining as $\mathbf{p} = \alpha_e \mathbf{E}_{in}$ and $\mathbf{m} = \alpha_m \mathbf{B}_{in}$.[29,116,117] It is noted that the definitions here are different from those in Ref.[45,46,118,119]. The exact dipole electric and magnetic polarizabilities can be expressed via conventional Mie scattering coefficients $a_1$ and $b_1$ as[45,120]

$$\alpha_e = \frac{3i\varepsilon a_1}{2k^3} \approx \frac{\alpha_e^{(0)}}{1-i\frac{2k^3}{3\varepsilon}\alpha_e^{(0)}}, \quad \alpha_m = \frac{3ib_1}{2\mu k^3} \approx \frac{\alpha_m^{(0)}}{1-i\frac{2\mu k^3}{3}\alpha_m^{(0)}}. \quad (3)$$

Here $k = \sqrt{\varepsilon\mu}\frac{\omega}{c}$; $\alpha_e^{(0)} = \varepsilon a^3 \frac{\varepsilon_p - \varepsilon}{\varepsilon_p + 2\varepsilon}$ and

$\alpha_m^{(0)} = \mu^{-1} a^3 \frac{\mu_p - \mu}{\mu_p + 2\mu} + O\left([ka]^5\right) = \frac{1}{k^3}\frac{(\varepsilon_p - \varepsilon)}{30\varepsilon}(ka)^5 \Big|_{\mu_p = \mu = 1}$ are the electrostatic limits of the

polarizabilities, $\varepsilon_p$ and $\mu_p$ are the electric permittivity and magnetic permeability of the particle, respectively. The last ones in Eq. (3) can be found via expending the scattering coefficients with respect to the small parameter ($ka$), higher terms can be found in Ref.[116,120].

Moreover, for nonmagnetic ($\mu_p = 1$) homogeneous particles,[45] $\alpha_e \sim (ka)^3$ and $\alpha_m \sim (ka)^5$. It means that in most cases only electric dipole approximation is already a sufficiently good approximation. The situation is different for the magnetic, anisotropic, bi-anisotropic, chiral, and more complex particles.[121-125]

Maxwell stress tensor in Eq. (1) contains the full field, which can be decomposed into incident and scattered fields as $\mathbf{E} = \mathbf{E}_0 + \mathbf{E}_{sca}$. Rigorously, the dipolar magneto-dielectric particle model shows an insightful analytic expression, which does not only take into account the forces from electric and magnetic dipoles but also the interactions between them. In the range of the dipolar approximation, the force can be expressed as[29,116,118,119]

$$\langle \mathbf{F} \rangle = \langle \mathbf{F}_e \rangle + \langle \mathbf{F}_m \rangle + \langle \mathbf{F}_{em} \rangle, \quad (4)$$

where $\langle \mathbf{F}_e \rangle$, $\langle \mathbf{F}_m \rangle$ and $\langle \mathbf{F}_{em} \rangle$ denote the electric dipole force, the magnetic dipole force, and the force from the interference between the electric and magnetic dipoles, respectively. Equation (4) can be further written as

$$\langle \mathbf{F} \rangle = \frac{1}{2}\mathrm{Re}\left\{ \underbrace{\mathbf{p}\cdot(\nabla)\mathbf{E}_{in}^*}_{\sim(ka)^3} + \underbrace{\mathbf{m}\cdot(\nabla)\mathbf{B}_{in}^*}_{\sim(ka)^5 \text{ for } \mu_p = \mu} - \underbrace{\frac{2k^4}{3}\sqrt{\frac{\mu}{\varepsilon}}(\mathbf{p}\times\mathbf{m}^*)}_{\sim(ka)^8 \text{ for } \mu_p = \mu} \right\}, \quad (5)$$

For arbitrary vectors **A** and **B**, $\mathbf{A}\cdot(\nabla)\mathbf{B} = \sum_{i=x,y,z} A_i \nabla B_i$.[126] The corresponding three terms of Eq. (5) represent the forces due to the induced electric dipole $\langle \mathbf{F}_e \rangle$, induced magnetic dipole $\langle \mathbf{F}_m \rangle$ and interference $\langle \mathbf{F}_{em} \rangle$ between them. Further on, the last term in Eq. (5) is responsible for a number of effects such as curl-spin related momenta transfer,[45,49] recoil force for Kerker-like effects, and gives important contributions to the pulling forces. Sometimes, usually even more multipoles should be considered.[30,127]

### C. Optical forces as a measure of field properties

Fundamentally, the optical force on any object comes from the difference in momenta between scattered and incident fields. To reveal this relation of the force on small dipole particles to incident field properties, Equation (5) can be rewritten in terms of well-known canonical momenta[109,111] and energy density as it was also recently done in acoustics[128]. The strategy of dealing with dipole electric $\langle \mathbf{F}_e \rangle$ and dipole magnetic $\langle \mathbf{F}_m \rangle$ terms is to achieve the allocation between different contributions related to real and imaginary parts of polarizability. The real part of the polarizability is responsible for gradient related forces or conservative contributions, on the other side, the imaginary part is directly connected with the extinction cross section $\sigma_{ext} = \sigma_{sca} + \sigma_{abs}$, i.e., with all scattering and absorption properties of the particle. As for the mix interference term $\mathbf{F}_{em}$ the most meaningful results are achieved by allocating contributions which are proportional to $\mathrm{Re}(\alpha_e \alpha_m^*)$ and $\mathrm{Im}(\alpha_e \alpha_m^*)$. Let us discuss each term.

The electric dipole force [$\mathbf{F}_e$ in Eq. (4)] can also be rewritten in terms of canonical momentum as[45,116,119]

$$\langle \mathbf{F}_e \rangle = \frac{4\pi}{\varepsilon} \mathrm{Re}(\alpha_e) \nabla \langle W_e \rangle + 2\frac{c}{n} \sigma_e^{ext} \langle \mathbf{p}_O^e \rangle, \qquad (6)$$

where $\langle W^e \rangle = \dfrac{\varepsilon |\mathbf{E}|^2}{16\pi}$ is the electric part of the time-averaged energy density,[45,49] $\langle \mathbf{p}_O^e \rangle$ is the electric contribution to the canonical (orbital) momentum $\langle \mathbf{p}_O \rangle$. The first and second terms on the right side of Eq. (6) correspond to the contributions of the electric part of the inhomogeneous energy distribution (optical gradient force) and the orbital momentum density, respectively.

The second term on the right side of Eq. (4) that represents the magnetic dipole force can also be expressed as[46,47,118]

$$\langle \mathbf{F}_m \rangle = 4\pi\mu \mathrm{Re}(\alpha_m)\nabla\langle W_m \rangle + 2\frac{c}{n}\sigma_m^{ext}\langle \mathbf{p}_O^m \rangle, \quad (7)$$

where $\langle W^m \rangle = \frac{\mu|\mathbf{H}|^2}{16\pi}$ is the magnetic part of the time-averaged energy density, $\langle \mathbf{p}_O^m \rangle$ is the magnetic contribution to the canonical (orbital) momentum $\langle \mathbf{p}_O \rangle$. Akin to $\langle \mathbf{F}_e \rangle$, the first term on the right side of Eq. (7) also corresponds to the optical gradient force; $\langle \mathbf{p}_O^m \rangle$ in the second term is the magnetic contribution to the orbital momentum density $\langle \mathbf{p}_O \rangle$. The cross sections are in dipole approximation relate to electric and magnetic polarizabilities as such

$$\sigma_{ext} = \sigma_e^{ext} + \sigma_m^{ext} = \frac{4\pi k}{\varepsilon}\mathrm{Im}(\alpha_e) + 4\pi k\mu\mathrm{Im}(\alpha_m), \quad (8)$$

$$\sigma_{sca} = \sigma_e^{sca} + \sigma_m^{sca} = \frac{8\pi}{3\varepsilon^2}k^4|\alpha_e|^2 + \frac{8\pi\mu^2}{3}k^4|\alpha_m|^2. \quad (9)$$

And $\sigma_{(e,m)}^{abs} = \sigma_{(e,m)}^{ext} - \sigma_{(e,m)}^{sca}$ being the asobtion cross section.

We note that the second term in eqs. (6-7) can be rewritten in terms of Pyonting vector and curl-spin contributions due to the non-uniform helicity as it is done in Refs.[116,129] using Poynting vector decomposion $\frac{1}{c^2}\langle \mathbf{S} \rangle = \frac{1}{n^2}\left(\mathbf{P}^O + \frac{1}{2}\nabla \times \mathbf{L}_S\right)$ [118,126]. Here we want to stress that the non-conservative part of the dipole force is colinear with the canonical linear momenta $\langle \mathbf{P}^O \rangle$ but not with the kinetic momentum density $\frac{1}{c^2}\langle \mathbf{S} \rangle$. To observe the curl-spin contribution one needs to consider higher order corrections which is present in $\langle \mathbf{F}_{em} \rangle$ [45,49].

The force $\langle \mathbf{F}_{em} \rangle$ from the interference between electric and magnetic dipoles can be expressed with terms of spin and orbital momenta as well as real and imaginary parts of *complex Poynting* vector as[47,118]

$$\langle \mathbf{F}_{em} \rangle = -\frac{8c\pi k^4}{3n}\left[\mathrm{Re}(\alpha_e\alpha_m^*)\frac{1}{n^2}\left(\langle \mathbf{p}_O \rangle + \frac{1}{2}\nabla \times \langle \mathbf{L}_S \rangle\right) + \mathrm{Im}(\alpha_e\alpha_m^*)\frac{1}{c^2}\langle \mathbf{S} \rangle^{Im}\right], \quad (10)$$

where $k$ is the wave number in the medium, $\langle \mathbf{L}_S \rangle$ is the spin angular momentum density of light, which can be described by the electric and magnetic contributions in a homogenous non-absorbing medium as[130]

$$\langle \mathbf{L}_S \rangle = \langle \mathbf{L}_S^e \rangle + \langle \mathbf{L}_S^m \rangle = \frac{1}{16\pi\omega} \text{Im} \left[ \varepsilon \mathbf{E}^* \times \mathbf{E} + \mu \mathbf{H}^* \times \mathbf{H} \right], \tag{11}$$

Meanwhile,

$$\langle \mathbf{p}_O \rangle = \langle \mathbf{p}_O^e \rangle + \langle \mathbf{p}_O^m \rangle = \frac{1}{16\pi\omega} \text{Im} \left[ \varepsilon \mathbf{E}^* \cdot (\nabla) \mathbf{E} + \mu \mathbf{H}^* \cdot (\nabla) \mathbf{H} \right]. \tag{12}$$

We can see from Eqs. (6) and (7), the forces from pure electric and magnetic dipoles are only related to the field gradient and orbital momentum of light, which are commonly known as the optical gradient force and orbital force, respectively. The force $\langle \mathbf{F}_{em} \rangle$ from the interference between electric and magnetic dipoles, though normally weak, can be used to investigate the lateral forces from the transverse angular momentum in some systems, such as two-wave interference field,[47] evanescence waves,[45,49] as well as spin-orbit interactions.[38,42,131]

For a vortex beam propagating in a free space with $\varepsilon = \mu = 1$ along the $z$-axis:[46,132-134]

$$\mathbf{E}(r,\theta,z) \simeq A(r,z) \frac{\bar{\mathbf{x}} + m\bar{\mathbf{y}}}{\sqrt{1+|m|^2}} \exp(ikz + i\ell\theta), \tag{13}$$

where the complex number $m$ determines the polarization state with $\sigma = \frac{2\,\text{Im}\,m}{1+|m|^2} \in [-1,1]$ being the helicity, the light with Hilbert factor $\exp(i\ell\theta)$ carries an OAM of $\ell\hbar$ ($\ell = 0, \pm 1, \pm 2, \ldots$) per photon.[15,132,135] Here, the small longitudinal $z$-components of the electric and magnetic fields are neglected, so "$\simeq$" is used in Eq. (13). In the paraxial approximation, the time-averaged energy density $\langle W \rangle$ and OAM of a monochromatic optical field in vacuum can be given as[46,132]

$$\langle W \rangle \simeq \frac{|A|^2}{8\pi}, \quad \langle \mathbf{L}_O \rangle \simeq \frac{\langle W \rangle}{\omega} \left( -rk\bar{\theta} + \ell\bar{\mathbf{z}} \right), \tag{14}$$

where $\langle W \rangle = \langle W^e \rangle + \langle W^m \rangle$. The orbital momentum density $\langle \mathbf{p}_O \rangle$ can be expressed as

$$\langle \mathbf{p}_O \rangle \simeq \frac{\langle W \rangle}{\omega} \left( k\bar{\mathbf{z}} + \frac{\ell}{r} \bar{\theta} \right). \tag{15}$$

Substituting Eq. (15) into Eqs. (6) and (7), we get the force from the electric and magnetic orbital momentum as

$$\langle F_{\mathbf{p}_o} \rangle \simeq \frac{2c}{\omega} \left( k\bar{\mathbf{z}} + \frac{\ell}{r} \bar{\theta} \right) \left[ \sigma_e^{\text{ext}} \langle W^e \rangle + \sigma_m^{\text{ext}} \langle W^m \rangle \right]. \tag{16}$$

The orbital force has been extensively used to rotate particles and cells in vortex beams, which according to Eq. (16) is related to the phase factor $\ell\bar{\theta}$. For the case of bigger interaction between higher multipoles takes place. As a consequence, the orbiting direction of particle could be unintuitive, i.e., have opposite sing relative to the azimuthal component of an OAM. It was theoretically predicted for the fiber configuration by F. L. Kien and A. Rauschenbeutel in 2013[136] and experimentally verified by G. Tkachenko at al. in S.N. Chormaic's group in 2020[137].

### D. Phase-dependent optical forces

Trapping particles in light beams is widely used in various experimental setups for the investigations of forces and testbed of extraordinary new optical phenomena.[34,138-140] For a weakly focused beam electric field where the phase varies spatially much stronger than the amplitude,

$$\mathbf{E}(\mathbf{r},t) = E_0(\mathbf{r})\mathbf{u}_{pol}e^{i\varphi(r)}, \tag{17}$$

where $E_0(\mathbf{r})$ is the slow varying amplitude, $\varphi(r)$ is the fast-varying phase, and $\mathbf{u}_{pol}$ is the polarization vector. Substituting Eq. (17) into the electric dipole force in Eq. (5) [the first term on the right side of Eq. (5)], it yields,

$$\langle \mathbf{F} \rangle = \frac{1}{4}\text{Re}(\alpha_e)\nabla E_0^2 + \frac{1}{2}\text{Im}(\alpha_e)E_0^2\nabla\varphi, \tag{18}$$

The first term on the right side of Eq. (18) represents the optical force from the intensity gradient, which is the same as the first term on the right side of Eq. (6), while the second term on the right side of Eq. (18) is the force from the phase gradient, which can be related to the scattering force on the particle. We note that the second term is a non-conservative part of the force. It has the form of phase gradient only since we have considered a weakly focused beam. The phase-gradient force has been used to rotate particles in a vortex beam,[41,117,132,141] or trap, bind and sort nanoparticles in line-shaped light beams.[21,140,142,143] The convenience of tailoring the amplitude and phase using metamaterials can endow more possibilities to harness amplitude- and phase-dependent optical forces for optical manipulations. Fundamentally, the optical forces increase linearly with the intensity as shown in Eq. (18), thus enhancing the intensity is an intuitive way to increase the optical force, while creating larger phase gradient also give rise to larger optical forces.

## E. Optical wavevectors-far field approximation and large particles

When the background medium is lossless with real permittivity $\varepsilon$ and real permeability $\mu$, the integration of Eq. (1) can be performed over any surface enclosing the obstacle, due to conservation of the electromagnetic momentum. Then the optical force can be valued over a spherical surface $S_\infty$ with a radius $R_S \to \infty$, which can be written as the integration of Poynting vectors,

$$\begin{aligned}\langle \mathbf{F} \rangle &= -\frac{n}{c}\int_{S_\infty}\left\{\langle \mathbf{S}\rangle - \langle \mathbf{S}^{(i)}\rangle\right\}dS = -\frac{n}{c}\int_{S_\infty}\left\{\langle \mathbf{S}^{(\text{mix})}\rangle + \langle \mathbf{S}^{(s)}\rangle\right\}dS\infty \\ &= -\frac{n}{8\pi}\int_{S_\infty}\left\{\text{Re}\{\mathbf{E}\times\mathbf{H}^*\} - \text{Re}\{\mathbf{E}^{(i)}\times\mathbf{H}^{(i)*}\}\right\}dS \\ &= -\frac{n}{8\pi}\int_{S_\infty}\left\{\text{Re}\{\mathbf{E}^{(i)}\times\mathbf{H}^{(s)*} + \mathbf{E}^{(s)}\times\mathbf{H}^{(i)*}\} + \text{Re}\{\mathbf{E}^{(s)}\times\mathbf{H}^{(s)*}\}\right\}dS\end{aligned} \quad , \qquad (19)$$

where $\langle \mathbf{S}\rangle$, $\langle \mathbf{S}^{(i)}\rangle$ and $\langle \mathbf{S}^{(s)}\rangle$ denote the total, incident and scattered time averaged Poynting vectors, respectively. $\langle \mathbf{S}^{(\text{mix})}\rangle$ is the mixed Poynting vector from the incident and scattered waves. $\mathbf{E}^{(i)}$ and $\mathbf{H}^{(i)}$ are the incident electric and magnetic fields, respectively. In some scenarios with strong electromagnetic resonances (e.g., multipoles), the incident wave becomes negligible compared with the scattered wave, wherein the plot of scattering Poynting vectors can be used to interpret the direction of the optical force.

When the particle is much larger than the wavelength ($ka \gg 1$), the optical force can be characterized by the geometrical optics, which uses the momentum exchange between different mediums to describe the light-matter interaction. The wavevector inevitably alters the direction after the reflection or refraction. According to Minkowski momentum of light, the recoil force exerted on the interface of two mediums can be given by[29,144,145]

$$\langle \mathbf{F}_{\text{ray},0}\rangle = \frac{n_i P_i}{c}\hat{\mathbf{r}}_i - \frac{n_i P_r}{c}\hat{\mathbf{r}}_r - \frac{n_t P_t}{c}\hat{\mathbf{r}}_t, \qquad (20)$$

where $P_i$, $P_r$ and $P_t$ denote light powers of the incident, reflected and refracted light rays, respectively; and $n_i$ and $n_t$ are refractive indices of the incident and refractive mediums, respectively. When a single light ray is reflected and refracted multiple times in a particle, the optical force on a sphere can be expressed as[144]

$$\langle \mathbf{F}_{\text{ray}} \rangle = \frac{n_i P_i}{c}\hat{\mathbf{r}}_i - \frac{n_i P_r}{c}\hat{\mathbf{r}}_{r,0} - \sum_{n=1}^{+\infty}\frac{n_t P_{t,n}}{c}\hat{\mathbf{r}}_{t,n}, \qquad (21)$$

where $\hat{\mathbf{r}}_i$, $\hat{\mathbf{r}}_{r,0}$ and $\hat{\mathbf{r}}_{t,n}$ are unit vectors of incident ray, the first reflected and $n$th refracted rays, respectively.

### F. Optical forces near surfaces

Manipulating particles near a surface is a common scenario in "meta-tweezers". Assuming the case of a dipole located in the vicinity of a solid substrate, the general expression for the optical force acting on a point dipole positioned at $\mathbf{r}_0$ can be written as[146-149]

$$\langle \mathbf{F} \rangle = \frac{1}{2}\text{Re}\left[\mathbf{p}\cdot(\nabla)\mathbf{E}_{\text{loc}}^*\right] = \langle \mathbf{F}_0 \rangle + \langle \mathbf{F}_{\text{recoil}} \rangle, \qquad (22)$$

where $\mathbf{E}_{\text{loc}}$ is the local electric field, and the self-consistent field $\mathbf{E}_{\text{loc}} = \mathbf{E}_0 + \mathbf{E}_D$ consists of two parts: the background field $\mathbf{E}_0$ acts on dipole and re-scattered field $\mathbf{E}_D$ from the dipole itself. The re-scattered field produced by a dipole can be written by the Green's function as $\mathbf{E}_D(\mathbf{r}) = 4\pi\mu\omega^2\hat{\mathbf{G}}(\mathbf{r},\mathbf{r}_0)\mathbf{p}$, which contains all the information about the nearby surface.[100,146-149] The Green's function has to be taken at its origin, i.e., at the position of the particle $\mathbf{r}_0$. The mechanism of the recoil force could be different. For example, particle under metallic surface can excite surface plasmon polariton (SPP) via near field interaction. In this scenario the recoil force will arise from conservation of momenta once SPP has directionality in oblique incident geometry.[147]

## III. METASURFACES AND METAMATERIALS FOR OPTICAL MANIPULATION

### A. Modulation of phase and amplitude

Metasurfaces are made of artificial subwavelength elements with small thickness which provide novel capabilities to manipulate electromagnetic waves. Here, we discuss some basic properties of metasurfaces employed for optical manipulation such as phase modulation, since it plays a key role in shaping optical fields, including their amplitude, polarization, wavevector, and spatial distribution. Basically, there are known several types of approaches to achieve phase modulation with metasurfaces, e.g. employing propagation phase,[150] resonant phase,[75]

Pancharatnam-Berry (PB) phase,[78] and exceptional topological phase.[151] To control the optical fields, the phase modulation must cover a full $2\pi$ range.

A metasurface consisting of either plasmonic or dielectric subwavelength elements can be designed to possess resonances that tailor both reflection and transmission of light. Since only $\pi$ phase change can be achieved for one simple resonance in accord with the Lorentz model, more complicated designs based on metal-dielectric-metal (MDM) structures[152] or a combination of electric and magnetic resonances[85] are usually used to engineer both reflection and transmission. Here, we consider MDM metasurface as an example. Based on the coupled mode theory, the reflection phase $\varphi_{\text{ref}}$ can be described as,[153]

$$\varphi_{\text{ref}} = \arg(r) = \text{angle}\big(\gamma^2 - (\omega - \omega_0)^2 - i\,2\gamma(\omega - \omega_0)\big). \tag{23}$$

Here $r$ is the reflection coefficient, and $\gamma$ and $\omega_0$ are the damping rate and resonance frequency related to the material and geometric size of metastructures. By rationally choosing the material and designing the geometric size of the metastructures, one can control the phase in a full $2\pi$ range from Eq. (23). However, we need to mention an elephant in the room: metallic structures are quite absorptive in the optical range which introduce its own constrains. The Huygens' phase combining both electric and magnetic resonances was proposed to increase the efficiency.[85] Another efficient phase modulation mechanism is to use propagation phase $\varphi_{\text{prop}}$ from a dielectric metasurface[150] given as,

$$\varphi_{\text{prop}} = n_{\text{eff}} k d \tag{24}$$

where $n_{\text{eff}}$ is the effective refractive index, $d$ is the thickness of a metastructure, and $k$ is the wavenumber of light. By adjusting the filling factor of the dielectric structure in each period, $n_{\text{eff}}$ can be controlled, such that a full $2\pi$ propagation phase can be realized.

However, the resonant phase and propagation phase usually exhibit strong dispersion that compromise the broadband operation of the optical devices. It is found that by simply rotating a birefringent meta-structure, a wavelength independent PB phase can be obtained. When the meta-structure is rotated with an angle of $\theta$, the rotated Jones matrix can be written as,[153]

$$\hat{J}(\theta) = \tfrac{1}{2}(s_u + s_v)\hat{I} + \tfrac{1}{2}(s_u - s_v) \times \big(e^{-i2\theta}\hat{\sigma}_+ + e^{i2\theta}\hat{\sigma}_-\big) \tag{25}$$

where the first term represents the helicity unconverted part, and the second term represents the helicity converted part. Here $s_u$ and $s_v$ are the complex reflection / transmission of the linearly polarized light passing through the metasurface along two principal axes ($\hat{u}$ and $\hat{v}$), $\hat{I}$ is the identity matrix. $\hat{\sigma}_{\pm}$ are two spin-flip operators that satisfy $\hat{\sigma}_{\pm}|\pm\rangle = 0$ and $\hat{\sigma}_{\pm}|\mp\rangle = |\pm\rangle$ ($|+\rangle$ represents left circular polarization (LCP) and $|-\rangle$ represents right circular polarization (RCP)). It can be seen from Eq. (25) that under the illumination of circular polarized (CP) light, the helicity converted part is imposed by a phase shift of $\pm 2\theta$ for the output light of $|\mp\rangle$, which is the PB phase or geometric phase. Therefore, the PB phase is only related to the rotation angle of the meta-structure, which is non-dispersive and can be used in broadband applications.

Recently, another phase modulation mechanism by encircling a singularity in a two-dimensional parameter space, which is called exceptional topological phase, is also introduced in the metasurface design.[151,154] In a complex plane, if a closed path encircles the origin, the winding number of the path around the origin is $w = \frac{1}{2\pi}\oint d\phi = 1$ (where $\phi$ is the phase of the complex number on the path). Therefore, the accumulated phase along the path is,

$$\Phi = \oint d\phi = 2\pi \tag{26}$$

It is seen that the encircling phase is always $2\pi$ which is protected by the origin.

By using the abovementioned phase modulation mechanisms, it can design a metalens for the amplitude enhancement at the focus point in the far field. The phase distribution of the metalens must obey the following equation,[82]

$$\varphi(r) = -\frac{2\pi}{\lambda}\left(\sqrt{r^2 + f^2} - f\right) \tag{27}$$

where $f$ is the focal length, $r$ is the radial position of the metalens, $\lambda$ is the operating wavelength.

The amplitude enhancement in the near field is usually realized by plasmonic and dielectric resonances to strongly localize the electromagnetic fields. In general, in lossy plasmonic resonant structures, the field enhancement can be expressed as,[86,155]

$$\frac{|E|^2}{|E_0|^2} \propto \frac{Q_{tot}^2}{V_{eff} Q_{rad}} \tag{28}$$

Here $E$ and $E_0$ are the localized and incident fields, respectively, $V_{eff}$ is the effective mode volume which is usually deeply subwavelength that guarantees a strong field enhancement at the resonance, $Q_{tot}$ is the total quality factor (Q factor) that satisfies $Q_{tot}^{-1} = Q_{rad}^{-1} + Q_{dis}^{-1}$ where $Q_{rad}$ and $Q_{dis}$ are the radiative and dissipative quality factors, respectively. Therefore, we have

$$\frac{|E|^2}{|E_0|^2} \propto \frac{Q_{\text{tot}}}{V_{\text{eff}}\left(1 + \frac{Q_{\text{rad}}}{Q_{\text{dis}}}\right)} \quad . \tag{29}$$

The field enhancement is related to the total Q factor $Q_{\text{tot}}$ and the ratio of the radiative and dissipative $Q$ factor $\frac{Q_{\text{rad}}}{Q_{\text{dis}}}$. By increasing $Q_{\text{tot}}$ or decreasing $\frac{Q_{\text{rad}}}{Q_{\text{dis}}}$, one can enhance the field substantially. For the regime of bound states in the continuum (BICs) without any losses, the value of $Q_{\text{tot}}$ can go to infinite. For this case, $Q_{\text{dis}}$ vanishes and $Q_{\text{rad}}$ goes to infinite ($Q_{\text{dis}} \ll Q_{\text{rad}}$), leading to a vanished field enhancement. In the opposite case when $Q_{\text{dis}} \gg Q_{\text{rad}}$, we have $\frac{Q_{\text{rad}}}{Q_{\text{dis}}} \approx 0$ and $Q_{\text{tot}} \approx Q_{\text{rad}}$. The field enhancement is also small being proportional to the radiative Q factor. Therefore, the maximum field enhancement is achieved for an intermediate state of comparable values of $Q_{\text{dis}}$ and $Q_{\text{rad}}$, that can be realized in the regime of the Fano-resonance[156] or quasi-BIC[157] resonance.

## B. Optical wavevectors control

A variation of the wavevector of a light beam can be realized through its interaction with a metasurface with an engineered phase gradient imprinted on arrays of subwavelength resonators. According to the generalized Snell's law of refraction, if a phase gradient of $\frac{d\varphi}{dx}$ is applied in the $x$ direction of an interface, the relationship between the angle $\theta_t$ of the transmitted beam and the angle $\theta_i$ of the incident bean can be presented as,[75]

$$\sin(\theta_t)n_t - \sin(\theta_i)n_i = \frac{1}{k_0}\frac{d\varphi}{dx} \tag{30}$$

where $n_t$ and $n_i$ are the refractive indices of the two media, $k_0$ is the wavenumber in the vacuum. By employing the fact that $\boldsymbol{k_i} = k_0 n_i \hat{\boldsymbol{r}}_i$, $\boldsymbol{k_o} = k_0 n_t \hat{\boldsymbol{r}}_o$ are the input and output wavevectors ($\hat{\boldsymbol{r}}_i$ and $\hat{\boldsymbol{r}}_o$ are the unit vectors along the input and output directions of the light propagation), $\boldsymbol{k_{ix}} = k_0 \sin(\theta_i) n_i \hat{x}$ and $\boldsymbol{k_{ox}} = k_0 \sin(\theta_t) n_t \hat{x}$ are the input and output wavevectors in the $x$-direction, we can simplify Eq. (33) to the form:

$$\Delta \boldsymbol{k} = \boldsymbol{k_o} - \boldsymbol{k_i} = \boldsymbol{k_{ix}} - \boldsymbol{k_{ox}} = \frac{d\varphi}{dx}\hat{x} \tag{31}$$

We notice that a change of the wavevector is simply proportional to the phase gradient accumulated along the interface, and this can be achieved by employing different mechanisms of the phase modulation as discussed in Sec. III A.

The wavevector can be manipulated in photonic crystals, hyperbolic metamaterials, and many other structures. For example, for hyperbolic metamaterials,[91,158-160] the parallel ($\parallel$) and pedicular ($\perp$) permittivity ($\varepsilon$) to the anisotropy axis in the hyperbolic structure is opposite, i.e., $\varepsilon_\parallel < 0$ and $\varepsilon_\perp > 0$ or $\varepsilon_\parallel > 0$ and $\varepsilon_\perp < 0$. The isofrequency surface is given by[91]

$$\frac{k_x^2 + k_y^2}{\varepsilon_\parallel} + \frac{k_z^2}{\varepsilon_\perp} = \left(\frac{\omega}{c}\right)^2, \tag{32}$$

where $k_x$, $k_y$ and $k_z$ are the $x$, $y$ and $z$ components of the wave vector, respectively. For a hyperbolic metamaterial with metal-dielectric multilayers, the effective permittivity $\varepsilon_\parallel$ and $\varepsilon_\perp$ can be expressed as,[160]

$$\varepsilon_\parallel = \frac{\varepsilon_d \varepsilon_m}{\varepsilon_d f + \varepsilon_m (1-f)}, \quad \varepsilon_\perp = \varepsilon_m f + \varepsilon_d (1-f) \tag{33}$$

where $f$ is the ratio of the thicknesses of the metallic and dielectric layers in one period, $\varepsilon_d$ and $\varepsilon_m$ are permittivity of dielectric and metal, respectively. Therefore, by adjusting the thickness ratio and material properties of the layers, one can achieve the wavevector control.

### C. Angular momentum control

Light can carry both SAM and OAM. SAM is associated with the circularly polarized (CP) light that carries $S = \sigma \hbar$ ($\sigma = \pm 1$) per photon of the angular momentum depending on the handedness of the CP light ($\sigma = 1$ represents LCP and $\sigma = -1$ represents RCP). OAM is associated with the optical phase profile of the beam in a plane orthogonal to the propagation direction. The electric field of a light beam carrying OAM per photon of $L = l\hbar$ (the topological charge $l$ is an integer) is given as,[161]

$$E(r,\vartheta) = E_0(r) e^{il\vartheta} \tag{34}$$

where $r$ and $\vartheta$ are the polar coordinates of the $xy$ plane. The total angular momentum is a sum of SAM and OAM as $J = (S + L)\hbar$ per photon. A specially designed metasurface can transform a plane wave into a wave carrying OAM with the topological charge $l$ by simply arranging the subwavelength elements of the metastructure in accord with a phase distribution

$$\varphi(r,\vartheta) = l\vartheta, \tag{35}$$

which can be realized as discussed in Sec. III A. Interaction of SAM and OAM can also be realized by using metasurfaces, usually called J-plates.[162] For example, a PB phase distribution of $\varphi(r,\vartheta) = l\vartheta$ can convert a CP wave carrying $S = \pm \hbar$ and $L = l_1$ into an OAM wave with $S = \mp \hbar$ and $L = l + l_1$.

### D. The use of other degrees of freedom

Any spatial distribution of the light field can be realized by using metasurface holograms, or metaholograms. According to the Fraunhofer diffraction, the electric field distribution in the image plane can be described as,[163]

$$E_{(x_i,y_i)} = \frac{e^{ikz}}{i\lambda z} e^{\frac{ik}{2z}(x_i^2+y_i^2)} \mathcal{F}\left(E_{(x_h,y_h)}\right) \tag{36}$$

where $(x_i,y_i)$ and $(x_i,y_i)$ are the coordinates at the image plane and hologram plane, respectively, $E_{(x_i,y_i)}$ and $E_{(x_h,y_h)}$ are the electric field distribution at the image plane and hologram plane, respectively, $z$ is the distance between the two planes, $\mathcal{F}(\cdot)$ is the operator of the Fourier transform, and $\lambda$ is the operating wavelength. Since the factor $\frac{e^{ikz}}{i\lambda z}$ can be neglected with constant $z$, and the quadratic phase factor $e^{\frac{ik}{2z}(x_i^2+y_i^2)}$ does not change the intensity profile, the diffracted field is proportional to a square of the Fourier transform of the field in the hologram plane. By applying the Gerchberg and Saxton algorithms,[164] given arbitrary $E_{(x_i,y_i)}$, it is possible to calculate the phase distribution of $E_{(x_h,y_h)}$ with unity amplitude. Following by the phase addressing mechanism described in Sec. III A, phase-only metaholograms can be realized to generate spatially distributed intensity patterns for massive particle manipulation.

Polarization is also an important degree of freedom of light. Polarisation engineering can be combined with metaholograms creating vectorial metaholograms employed for the generation of holographic images with arbitrary polarization distributions.[165] Recently, high-order states of polarization of vector vortex beams that combines both the polarization and OAM have been proposed,[166] being described as,

$$|\psi_l\rangle = \psi_R^l|R_l\rangle + \psi_L^l|L_l\rangle \tag{37}$$

where $\psi_R^l$ and $\psi_L^l$ are the complex amplitudes of two orthonormal CP basis with OAM of $l$ as,

$$|R_l\rangle = e^{-il\vartheta}|R\rangle \tag{38}$$

$$|L_l\rangle = e^{il\vartheta}|L\rangle \tag{39}$$

It is also called a high-order Poincaré sphere beam since it can be represented on a high-order Poincaré sphere with high-order Stokes parameters. High-capacity and multiplexing of high-order Poincaré sphere beam using metasurfaces have also been proposed,[167,168] showing multifunctionality of metasurface applications.

Nonlinear optical response of metamaterials and metasurfaces provides a new degree of freedom for manipulating optical fields, which not only alters the output frequency, but also introduces new properties that are not observed in the linear regime.[74,169-171] For example, the geometric phase in the nonlinear regime works in co-polarization as well and it is highly dependent on the rotational symmetry of metastructures.[77,172-174] Metamaterials and metasurfaces also provide a great platform for quantum physics,[175,176] showing nonclassical phenomena that may find promising applications in single- or multi-photon optical tweezers. Apart from the optics, metamaterials and metasurfaces can be used for the manipulation of heat and acoustic waves,[177-181] which may open new opportunities for novel optical tweezer designs.

## IV. METAMATERIALS FOR OPTICAL MANIPULATION
### A. Shaping amplitude and phase

Optical trapping using metalens facilities implementing a planar platform for the compact chip-scale manipulation of particles.[182-186] The engineering of Pancharatnam–Berry phase enables the polarization-sensitive particle manipulation (see also Section III).[187,188] The adjustment of the position of metalens offers a three-dimensional particle manipulation. The metalens can also be used to trap and rotate birefringent nematic liquid crystal droplets by the anisotropic polarization in the light field.[189] The optical levitation, which joins the fields of optomechanics and optical trapping is currently under the spotlight, is a powerful technique for fundamental science, and now entering into the quantum regime,[190-192] enabling applications such as quantum state transfer, teleportation and entanglement.[193-195] On-chip optical levitation with a metalens minimize the system to make it portable, which can be used for numerous precision measurements and sensing applications.[196-199] The propagation-phase dielectric metasurface, composed of GaN circular nanopillars, was reported to generate a polarization-independent vertically accelerated two-dimensional Airy beam in the visible region,[28] allowing guiding and clearing particles in three dimensions, as shown in Fig. 2b. The utilization of metasurface replaces other conventional ways in generating the Airy beam, such as SLM, deformable mirror, digital micro-mirror device, and asymmetric nonlinear photonic crystals, which suffers from bulky size, redundant diffraction orders, and low transmission efficiency.[26,200] It can be envisioned that other non-diffraction and self-healing beams (e.g., Bessel beam) that have already been realized using metasurfaces can also be used to three-dimensionally manipulate particles.[201] The metalens can be further integrated into an optical fibre for flexible optical trapping,[202] paving the new way for increasing the transmission efficiency and mitigating the chromatic aberration, which are highly related to optical manipulations. This meta-fibre can potentially work as a fibre-optic endoscope to directly manipulate tiny objects in human beings.[203-205]

Strong field confinement is an intuitive way to enlarge the gradient part of optical forces (first terms in eqs. (6-7)). Plasmonic metamaterials overcome Abbe's diffraction limit and create electric fields with large gradients, facilitating high-efficient trapping of nanoparticles.[206-210] Recently, Kotsifaki et al. demonstrated the Fano resonance-assisted plasmonic optical tweezers to trap 20 nm polystyrene nanoparticles. The scaled trapping efficiency is $8.65 \text{ fN} \cdot \text{nm}^{-1} \cdot \text{mW}^{-1}$ for 20 nm polystyrene particles, being ~24 times larger than that for a coaxial nanoaperture cavity, which is promising for the on-chip trapping with ultra-low laser power.[211] Ndukaife et al. used the thermoplasmonics nanohole metasurface which induces thermal gradient and generates the electrothermoplasmonic flow to achieve the high-resolution and large-ensemble nanoparticle trapping.[212] This multidisciplinary technique that combines the electric and thermofluidic fields can manipulate nanoparticles with relatively lower temperate increment, thus more friendly to biological specimen.[213,214] More examples to enhance the electric field by strong electromagnetic resonances in metamaterials can be found in nanowire pairs,[215] twisted split-ring-resonators,[216] planar nanorods,[217] silicon nanotrimer,[218] cloaking,[219] plasmonic bowtie nanoantenna,[220-222] etc.

Recently, BICs, which were first proposed by von Neumann and Wigner in quantum potential wells, have aroused great attention in photonics.[223] One of the reasons of immense interest to BIC physics in photonic structures is the unprecedentedly almost infinite quality factors that gives rise to the perfect confinement of light,[224-226] consequently strong enhancement of near-fields and optical forces in some particular scenarios.[227] The perfect confinement of light in nanocavities by BIC can also result in a negligible leakage of light out of nanostructures, which generates a weak trapping force.[228] Meanwhile, structures that support

BICs have a great application perspective in field of optomechanical coupling[229] and in optomechanical modulation.[230] The true BIC has an infinite Q-factor but at the same time it is getting impossible to excite this mode due to the reciprocity principle. Moreover, for any open resonator true BIC does not exist (non-existence theorem[223]) with some very specific exemption utilizing epsilon-near-zero materials[231,232] and zero-coupling between shear waves in solids and longitudinal waves in host media in acoustics.[233]

So in practice, BICs are realized with high but finite quality factors due to structural losses and imperfections, and they are usually termed "quasi-BICs" or "supercavity modes" or "qBIC", and some of the quasi-BIC applications have been demonstrated for lasing,[225,234] sensing,[235] sensing[236] and other effects.[237] Yang et al. used the coupled nanopillars to generate the quasi-BICs, which enhanced the electromagnetic field by an order of magnitude higher than plasmonic systems.[238] The large array of dielectric resonators generates multiple hotspots for high-throughput trapping nanoparticles, providing a new approach to realize the low-power optical trapping. Apart from particles, the optical gradient force between two waveguides can be enhanced using transformation optics, by attaching the waveguide with the single-negative metamaterial working as the annihilating medium.[239] The key players here are the gradient forces since qBIC modes have extremely low radiation losses.

As we can see, metamaterials show unparalleled advantages in shaping the amplitude and phase of light field, which perfectly coincides with the principle of optical tweezers (see Sec. II D), allowing multi-dimensional optical trapping of nanoparticles, and pushing the limit of trapping towards smaller, lower power, more robust and more compact. Metamaterials allow the precise manipulation of amplitude and phase at the sub-wavelength scale, being more advanced than the way with refractive optics, which uses the SLM, deformable mirror, digital micro-mirror device, etc. Moreover, the much more compact configurations using metamaterials can offer more opportunities for on-chip optical manipulations, and greatly reduce the cost, facilitating the development of affordable devices for biomedical applications, such as trapping, sensing and tumour targeting.

### B. Engineering momentum topology

Metamaterials also provide great flexibility in engineering the momentum topology of light. Among them, one of the unusual classes is the hyperbolic metamaterial as described in Sec. III B, which displays hyperbolic dispersion, determined by their effective electromagnetic tensors.[91,159] The existence of hyperbolic metamaterial reshapes the scattered momentum in the

*x*-direction from $k_x^0$ to $k_x^{\text{hyp}}$, thus imposing a net increased momentum (Fig. 3a), which in return, pulls the particle[160,240-242] (Fig. 3b). It is noted that, when the metallic particle (sphere or elliptical) is placed above the metal-dielectric hyperbolic metamaterial, the excitation of directional SPPs also contribute to the optical pulling force along with the momentum topology.[33,147,149]

Two particles can bunch together in the hyperbolic metamaterial to bind with a distance controlled by optical modes,[146] which can be much longer than the classical optical binding as shown in Fig. 3c. Giant optical lateral forces can also be realized on Rayleigh particles placed near hyperbolic and extremely anisotropic metasurfaces.[148] This new binding mechanism may open an avenue for enhanced binding forces in biomedical applications, many-body interactions, just to name a few. Aside from enhanced and reversed optical forces, the enhanced optical torque on a quantum emitter placed inside a hyperbolic metamaterial can align it in a well-defined direction,[243] as shown in Fig. 3d. The self-induced torque can be of several pN·nm, potentially being important for biological applications, such as DNA folding. Recently, Qiu, Ding and their team proposed an ingenious mechanism to obtain the optical pulling force by transiting the light momentum from the usual convex to a concave shape using a photonic crystal meta-structure,[40] as shown in Fig. 3e. The triangle (Fig. 3f) or ellipse can effectively reflect light forward, generating a large optical pulling force, which however, exhibits in a wide range of geometric parameters. The hyperbolic and concave shapes of momentum vectors are just two examples demonstrated so far, more engineered momentum space in metamaterials can be expected for intriguing optical manipulations. For instance, arbitrarily control of the wavevectors can move particles along a three-dimensional complex trajectory, push and pull back the particle depending for different utilities.

### C. Spatiotemporal manipulation

The "active metamaterials" with superb tunability of light in the degrees of freedom of time and space have attracted much attention in recent years.[244-246] The tuning unit can be micromachined,[95,247,248] liquid pumped,[83,249,250] electrically powered,[73,92,251] thermal controlled by liquid crystal,[252] etc. The versatile tunability of metamaterials enables the optical manipulations in real time. The focusing points of a plasmonic metalens can be dynamically tuned with different polarizations (Fig. 4a), which could enable trapping of particle in different layers of microchannel,[253] providing numerous opportunities in compact multifunctional optofluidic manipulations, such as transporting and sorting.[19,20] A metalens tweezers can create shiftable thin sheets across the microchannel, which can be used for the optofluidic sorter in the microchannel.[254] Recently, Danesh et al. took advantage of graphene's Dirac plasmon for its extreme confinement and tunability, to build up a monolayer conveyor with movable potential wells for transporting sub-1 nm nanoparticles,[255] as shown in Fig. 4b. The theoretical scheme has great potential in the trapping, transporting, and sorting sub-10 nm nanoparticles, being more advanced than current cutting-edge mechanisms.[20,21,65,256]

Due to recent advances in nanofabrication technology, meta-structures can be peeled off from the substrate and become the active meta-robots operated by light. Those meta-robots take full advantages of interactions of light and nanostructure to serve as a new class of micro/nanomechanical devices. A plasmonic linear nanomotor proposed by Tanaka et al. uses the optical lateral force from the directional side scattering by the meta-nanostructure,[257] as shown in Fig. 4c. The nanomotor can move with a resolution beyond the diffraction limit operated by linearly polarized light beams, as it does not depend on light gradient but only on polarizations of light. With a similar principle but more elegant control both in translation and rotation, Andren et al. reported a microscopic meta-vehicles which can achieve a complex trajectory via dynamically controlling the force and torque from different polarizations of light[258] (Fig. 4c). The meta-vehicle can be aligned along the input polarization of linearly polarized beam and be rotated by circularly polarizations with rotating directions coinciding with the helicity of light. This meta-vehicle could provide tremendous biomedical applications, such as cargo transporting, tumour targeting,[259] etc.

Interestingly, Ilic and Atwater recently developed a unique "light sail" using meta-nanostructures which can be self-stabilized under the levitation and propulsion of light.[260] This microdevice elegantly controls the scattering light by engineering the scattered phase to create the spatial restoring force and torque automatically maintaining the "sail" stably inside the light beam, as shown in Fig. 4e. This technique does not require highly focused beam to trap particle inside, whereas the propulsion force in a collimated beam can guide the device along a long distance, being potentially feasible for the space travel. This innovation inspires further utilizations of metasurfaces for specially functionalized and audacious applications that conventional optical tweezers cannot achieve. By choosing suitable wavelengths, the metamaterial powered by light could have an optical "adhesion" force near a surface overcoming the radiation pressure, mimicking the Gecko toes sticking to the wall.[261] Dynamically tuning of the wavelengths or controlling the "on" and "off" of laser, the forces can be switched between the "adhesion" and propulsion forces, or between the "adhesion" and "zero" force, making the walking on a wall possible.

### D. Angular momentum manipulation

Angular momentum, an intrinsic property of light, consisting of the spin and orbital angular momenta, have been widely deployed to manipulate particles in classical optical tweezers, realizing particle spin/rotation,[14-16,262-264] optical lateral force,[34,45-49] etc. Metasurfaces facilitate to miniaturize the bulky three-dimensional components, such as microscope objectives and spatial light modulators for the realization of the on-chip optical spanner,[141,265] as shown in Fig. 5a. The optical torque on the particle can be increased with the increment of topological charge of the metalens. And the radius of the vortex ring can be controlled by selecting appropriate focal lengths and topological charges, providing a great degree of flexibility for the optical spanner, micro-motors, and other on-chip applications. The counterintuitive "left-hand" optical torque can be demonstrated using an inhomogeneous and anisotropic transparent macroscopic medium, which is the form-birefringent nanostructured glass plate,[266] as shown in Fig. 5b. The idea is to couple the incident SAM to OAM (spin-orbit interaction) using a phase plate to achieve the negative optical torque.

Meanwhile, the microscopic or macroscopic meta-rotor can be fabricated by micromachining of four phase plates with certain orientational gradient and being assembled together on the water surface,[267] or using a top-down approach with e-beam lithography exposure two times and subsequent etching the sacrificial substrate to lift off the meta-nanostructure.[258] By designing a one-dimensional geometric phase profile, Magallanes and Brasselet demonstrated the bilateral movement of a microscopic optical element illuminating with different helicity of light by spin-orbit interactions,[267] as show in Fig. 5c. Opposite spins of light will be scattered to two different sides after passing through the periodic phase grating, in return, resulting in opposite lateral forces.

Though not many phenomena about spin or angular momentum have been demonstrated so far, mostly focusing on the rotate of the particle or meta-plate, there are plenty of room for further experimental exploration, for instance, transverse spin, spin-orbit interactions, etc, to unveil the underlying physics of light-matter interactions.

### V. CONCLUSION AND OUTLOOK

Metamaterials, which are artificial materials containing sub-wavelength structures array, modify the permittivity and permeability to achieve numerous exotic characters beyond the nature, making them competent in freely controlling the dispersion, refraction, and reflection of electromagnetic waves. Their abilities to tailor versatile degrees of freedom of light beam make metamaterials a paradigm of particle manipulation. By engineering the metastructures using different phase modulation mechanisms, such as resonant phase, propagation phase and Pancharatnam–Berry phase, the metalens can trap, levitate and rotate particles on an integrated chip, and easily control the trapping positions in three dimensions; The strong confinement of light field by Fano resonance, anapole, multipoles and quasi-BIC could push the trapping limit towards smaller size, large quantity and high efficiency. Many of functions of particle manipulations can be realized using metamaterials, including pulling, lateral mobilizing and binding in momentum topology metastructures, sorting using thermoplasmonics meta-plate, and conveying using monolayer or conveyor belt.[268,269]

The recent emergence of meta-robot due to the fast development of nanofabrication technology has opened another realm for optical manipulations. Including the fabrication of plasmonic[257] and dielectric[258] meta-robots. Both methods require two times of e-beam lithography to create the nanostructure and the boundary of meta-robot. The major difference of them is the first step of lithography to pattern the silicon or metallic nanostructures. Instead of manipulating particles, the meta-structure itself takes the full advantage of the interactions with light, and moves in an extraordinary way. Examples include the self-stabilizing in a propulsion beam, spinning following the SAM, vehicles moving in straight and complex trajectories, lateral shifting by spin-orbit interactions, etc.

As the burgeoning development of metamaterials with the emerging new optical physics, the optical manipulations with metamaterials are expected to evolve. New ideas should come out shortly, just to name a few, the exceptional points in metamaterials offer a new mechanism to use light with a high sensitivity,[151,270-272] which may be utilized for the high-sensitive sorting of nanoparticles; the fascinating transvers spin in structured light beams (e.g., standing wave) and evanescence waves could generate optical lateral forces, which can also be realized in metasurfaces to enable potential bidirectional optical sorting and conveying; the meta-hologram could be used to trap and sort massive particles, and assembly particles in three dimensions;[273,274] the combination of different degrees of freedom, such as hologram and polarization,[153,165,275] hologram and OAM, enriches the optical manipulation and makes optical tweezers multifunctional;[150,276] the meta-robots designed with particular optical properties[277] and responses could enable enormous applications for biomedical science, such as drug delivery, molecule interactions. We envision that, with metamaterials, optical tweezers will become a more versatile and powerful tool in biophysics, as well as an ideal testbed for new ideas in emerging optics.

## ACKNOWLEDGMENTS

Y.S. acknowledges Natural Science Foundation of Shanghai, Grant No. 22ZR1432400. Q.S. acknowledges a start-up funding from the Tsinghua Shenzhen International Graduate School (SIGS), Tsinghua University, No. 01030100006. A.Q.L. acknowledges the Singapore Ministry of Education (MOE) Tier 3 grant (MOE2017-T3-1-001). Y.K. acknowledges a support from the Australian Research Council (grants DP200101168 and DP210101292) and the Strategic Fund of the Australian National University.

## AUTHOR DECLARATIONS

**Conflict of Interest**

The authors have no conflict of interest.

## DATA AVAILABILITY

The data that support the findings of this study are available within the article.

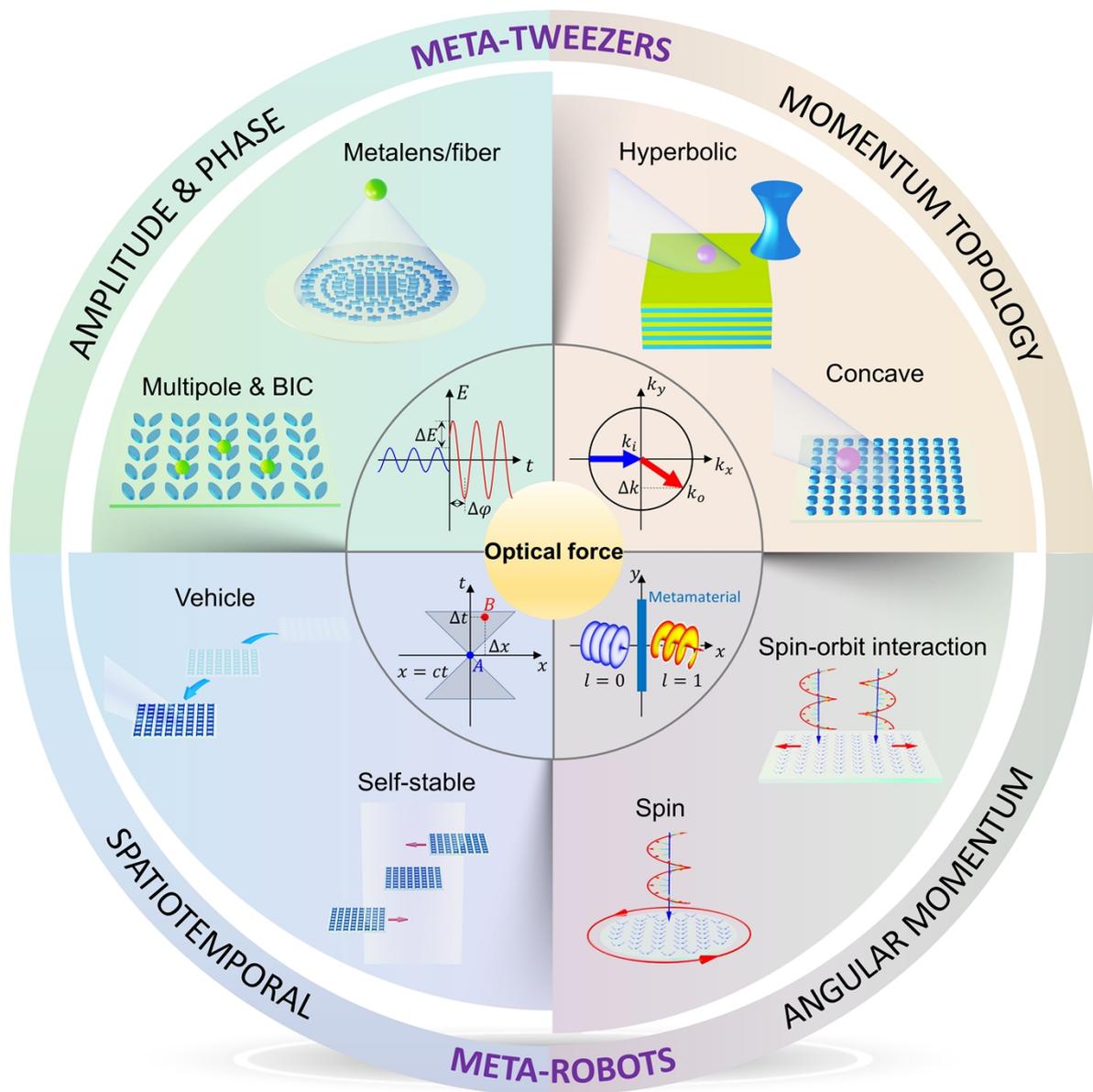

Fig. 1. Meta-tweezers for multidimensional manipulation of light. Schematic illustration of "meta-tweezers" from four properties of light wave: amplitude & phase, momentum topology, spatiotemporal and angular momentum, to realize the multifunctional manipulations of particles and active meta-structures (meta-robots).

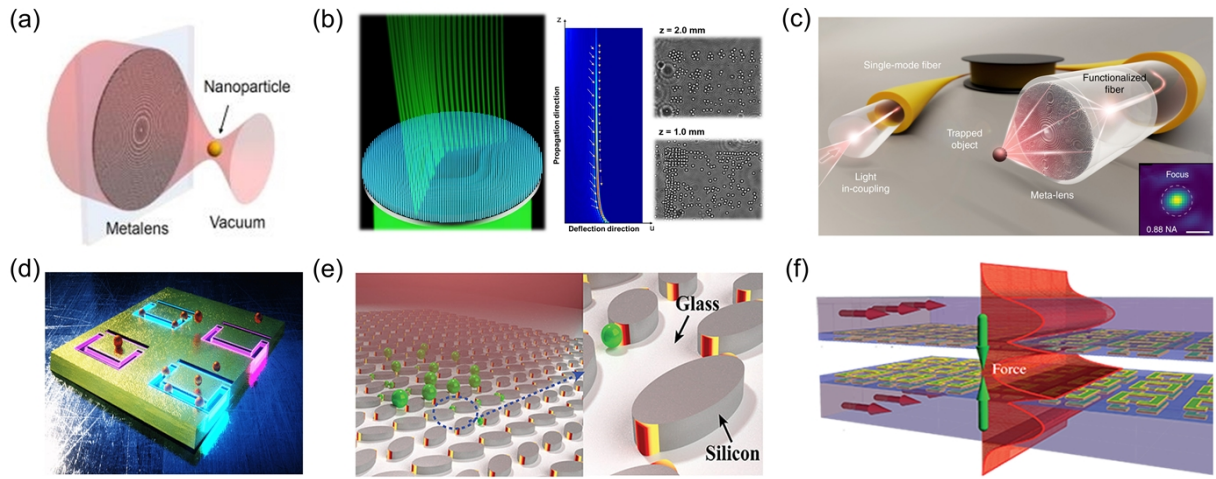

Fig. 2. Shaping amplitude and phase of light for meta-tweezers. (a) On-chip levitation of particles using a metalens. Reproduced with permission.[196] Copyright 2021, Optical Society of America. (b) Cubic-phase Airy beam for three-dimensional manipulation of particles. Reproduced with permission.[28] Copyright 2021, MDPI. (c) Metalens-fiber for a more portable and flexible optical trapping. Reproduced with permission.[202] Copyright 2021, Nature Publishing Group. (d) Fano resonance-assisted "meta-tweezers" for a high-efficiency trapping of 20 nm nanoparticles. Reproduced with permission.[211] Copyright 2020, American Chemical Society. (e) Exploring quasi-BIC for optical trapping. Reproduced with permission.[238] Copyright 2021, American Chemical Society. (f) Enhancing optical gradient force using transformation optics between meta-structures. Reproduced with permission.[261] Copyright 2012, American Physical Society.

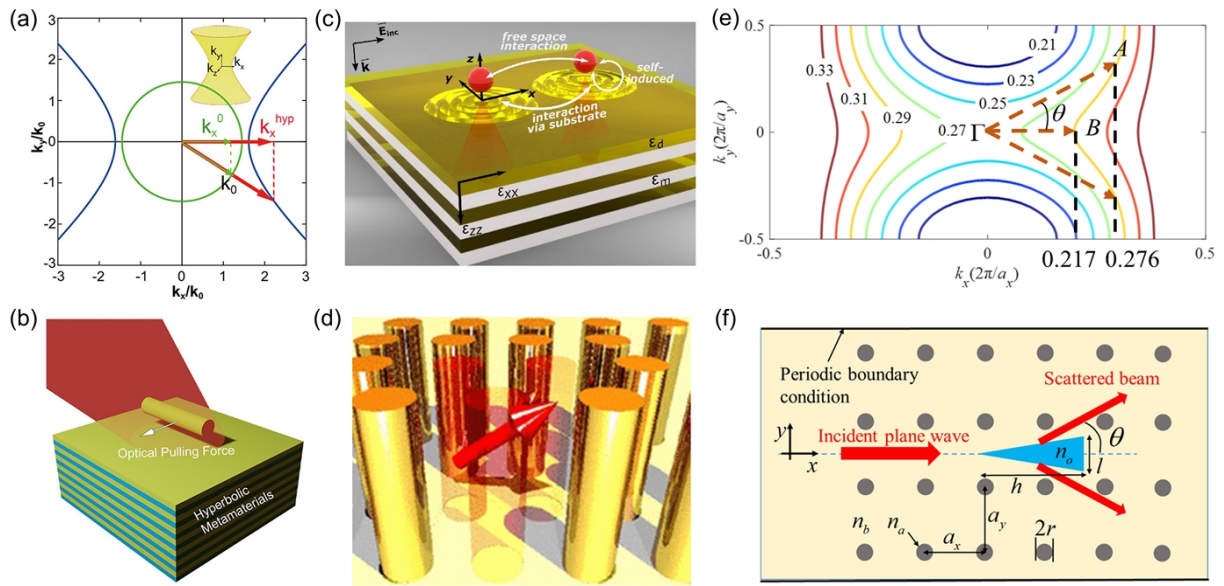

Fig. 3. Engineering momentum topology for optical manipulation. (a) Isofrequency two dimensional surfaces of waves in water (green) and hyperbolic metamaterials (blue). The momentum vector in red surpasses the green one in the +*x*-direction, resulting in an optical pulling force. Reproduced with permission.[160] Copyright 2021, American Chemical Society. (b) Illustration of optical pulling force above a hyperbolic metamaterial with metal-dielectric multilayers. Reproduced with permission.[160] Copyright 2021, American Chemical Society. (c) Nanoscale tunable optical binding above a hyperbolic metamaterial.[146] Copyright 2020, American Chemical Society. (d) Illustration of enhanced optical torque on a quantum emitter placed inside a hyperbolic metamaterial.[243] Copyright 2013, American Physical Society. (e) A special concave-shaped momentum space for optical pulling. Reproduced with permission.[40] Copyright 2020, American Physical Society. (f) Particle inside the nanostructures generates a dominant forward scattering. Reproduced with permission.[40] Copyright 2013, American Physical Society.

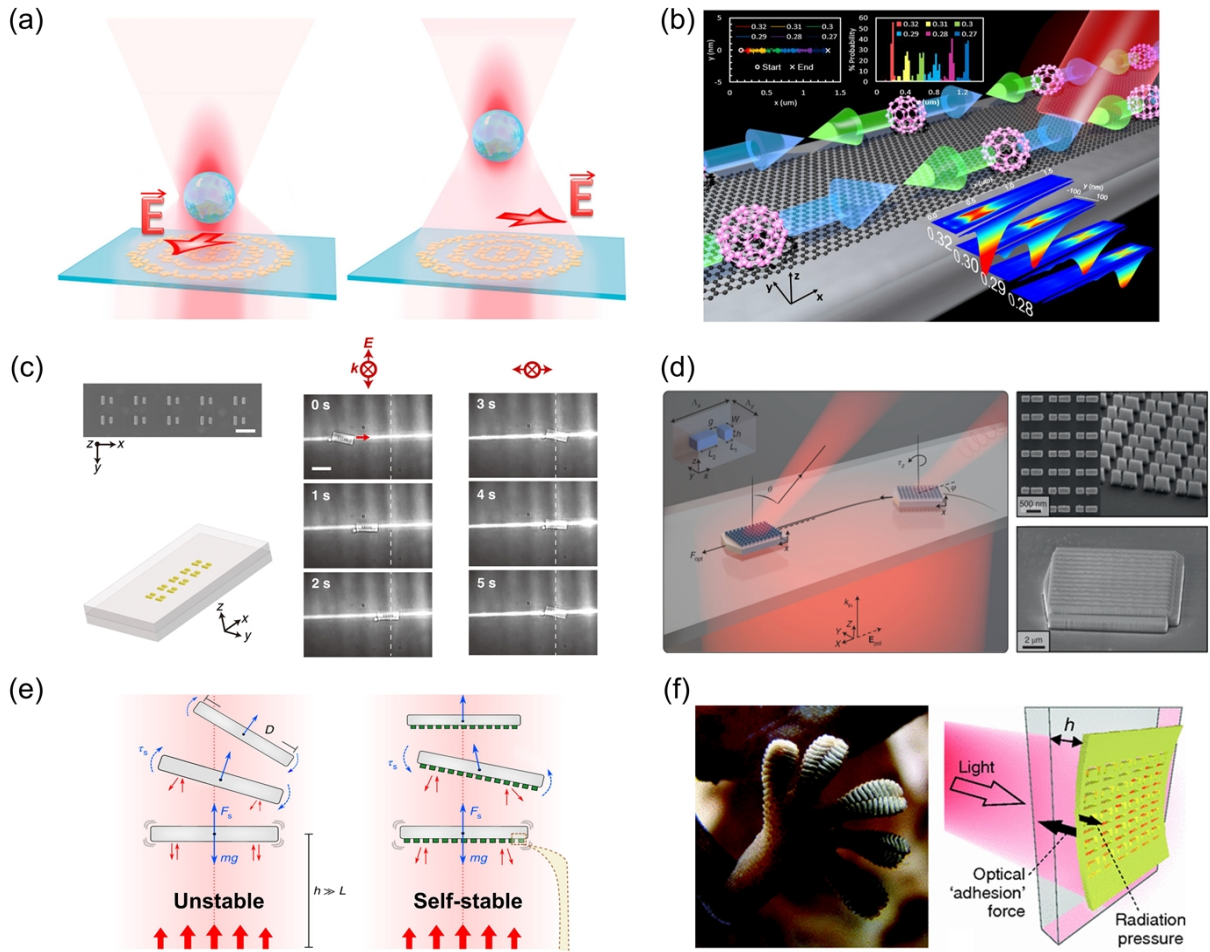

Fig. 4. Dynamic control of optical forces and potentials. (a) Tuning of focusing depth by different polarizations for optical trapping in a metasurface. Reproduced with permission.[253] Copyright 2018, American Chemical Society. (b) A monolayer conveyor by graphene's Dirac plasmon for sub-1 nm nanoparticle conveying.[255] Copyright 2020, Wiley-VCH. (c) Plasmonic meta-robot powered by a linearly polarized beam moving in a linear way.[257] Copyright 2020, American Association for the Advancement of Science. (d) Complex-trajectory meta-robot powered by the synergy of different polarizations of light.[258] Copyright 2021, Nature Publishing Group. (e) Self-stabilizing "light sail" under the levitation and propulsion of light. Reproduced with permission.[260] Copyright 2019, Nature Publishing Group. (f) Optical "adhesion" force and propulsion force on a metamaterial near a wall. Reproduced with permission.[261] Copyright 2012, American Physical Society.

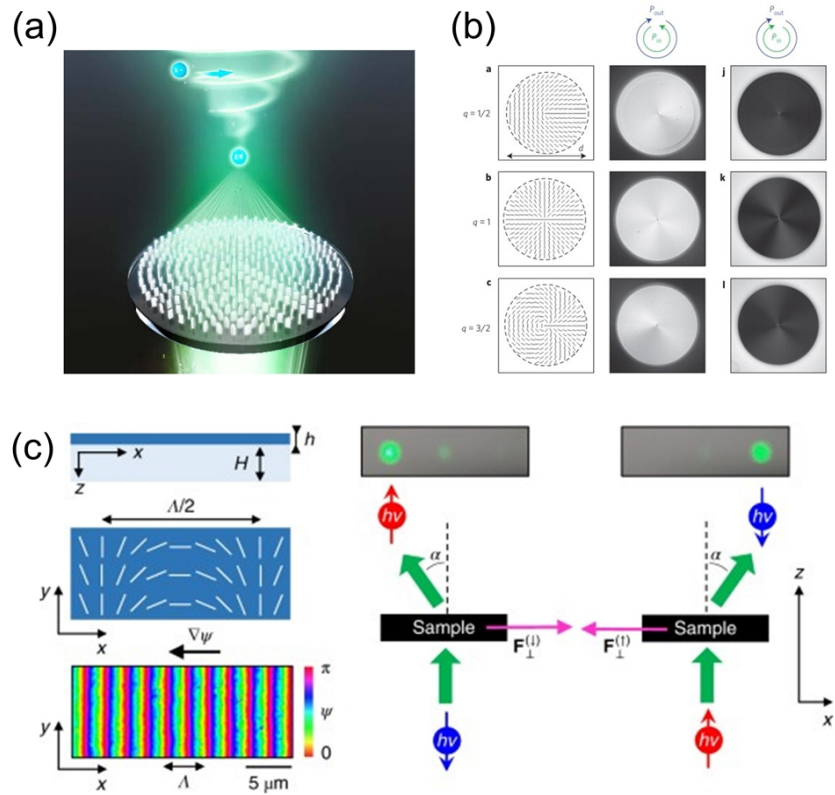

Fig. 5. Angular momentum of light employed for meta-tweezers. (a) On-chip vortex beam (spanner) based on geometric and dynamic phases. Reproduced with permission.[265] Copyright 2021, OSA and Chinese Laser Press. (b) Counterintuitive "left-hand" optical torque by the form-birefringent nanostructured glass plate.[266] Copyright 2020, Nature Publishing Group. (c) Bidirectional optical motor with spin-orbit interactions by different helicity of light.[267] Copyright 2018, Nature Publishing Group.